\documentclass[11pt]{article}
\usepackage{color}\usepackage{dirtytalk}
\usepackage{amsmath}

\def\whitebox{{\hbox{\hskip 1pt
        \vrule height 6pt depth 1.5pt
        \lower 1.5pt\vbox to 7.5pt{\hrule width
                  3.2pt\vfill\hrule width 3.2pt}%
        \vrule height 6pt depth 1.5pt
        \hskip 1pt } }}
\def\qed{\ifhmode\allowbreak\else\nobreak\fi\hfill\quad\nobreak
     \whitebox\medbreak}

\newcommand{\bc}{{\bf c}}

\newcommand{\remove}[1]{}

\newcommand{\proof}{\noindent{\bf Proof. }}

\newcommand{\F}{{\bf F}}

\newcommand{\B}{{\cal B}}

\newcommand{\Mt}{{\tilde M}}

\newtheorem{theorem}{Theorem}

\newtheorem{definition}{Definition}

\title{Efficient Conflict Resolution in Multiple Access Channels Supporting Simultaneous Successful Transmissions }
\author{\sc Annalisa  De Bonis\\
Universit\`a di Salerno,
84084 Fisciano (SA), Italy\\
email: \texttt{debonis@dia.unisa.it}} 

\date{}
\begin{document}
\maketitle
\begin{abstract}
We consider the {\em Conflict Resolution Problem} in the context of  a  multiple-access system in which several stations can transmit their messages simultaneously to the channel. We assume that there are $n$ stations and that at most $k$, $k\leq n$, stations are {\em active} at the same time, i.e, are willing to transmit a message over the channel. If in a certain instant at most $d$, $d\leq k$,  active stations transmit to the channel then their messages are successfully transmitted, whereas if more than $d$ active stations transmit simultaneously then  their messages are lost. In this latter case we say that a  {\em conflict} occurs. The present paper investigates  {\em non-adaptive} conflict resolution algorithms  working under the assumption that active stations receive a {\em feedback} from the channel that informs them on whether their messages have been successfully transmitted. 
If a station becomes aware that its message has been correctly sent over the channel then it becomes immediately {\em inactive}, that is, stops transmitting. The measure to optimize is the number of time slots needed to solve conflicts among all active stations. 
The fundamental question is whether this measure decreases linearly with the number $d$ of messages that can be simultaneously transmitted with success.
 In this paper we give a positive answer to this question by providing a conflict resolution algorithm that uses a $1/d$ ratio of the number of time slots used by the optimal conflict resolution algorithm for the particular case $d=1$ \cite{kog}. Moreover, we  derive a lower bound on the number of 
time slots needed to solve conflicts non-adaptively 
which is within a $\log (k/d)$ factor from the upper bound. To the aim of proving these results, we introduce a new  combinatorial structure that consists in a generalization of Koml\'os and Greenberg codes \cite{kog}. Constructions of these new codes are obtained via a new generalization of selectors \cite{dgv}, whereas the non-existential result is implied by a non-existential result for a new generalization of the locally thin families of \cite{afk,csurus}. We believe that the combinatorial structures introduced in this paper and the related results may be of independent interest. \end{abstract}

\section{Introduction}
Conflict resolution is a fundamental problem in multiple-access communication and has been widely investigated in the literature both for its practical implications and for the many theoretical challenges it poses \cite{Chl}. Commonly, this problem is studied under the assumption of the so called {\em collision model} in which simultaneous transmission attempts by two or more stations result in the destruction of all messages.  However, as already observed in \cite{gvs} and more recently in \cite{chan_berg}, this restrictive  multiaccess model does not capture the features of many important multiuser communication systems in which several messages can be successfully transmitted at the same time. Examples of these communication systems include  Code Division Multiple Access (CDMA)
systems  in  which several stations  share the same frequency band, and Multiple-Input Multiple-Output (MIMO) systems,  that enhance the capacity of a radio link by using multiple  antennas at the transmitter and the receiver. These systems  are largely used in the 
 phone standards, satellite communication systems, and in wireless communication networks. Given the growing  relevance of these systems in modern communication technologies, it is crucial to consider multiple-access models that better capture the phenomenon occurring when  multiple packets can be sent simultaneously over the channel. The following quotation from \cite{chan_berg} well emphasizes the importance of these communication models: 
 \say{Traditionally, practical design and theoretical analysis
of random multiple access protocols have assumed
the classical collision channel model --- namely, a tran\-smitted
packet is considered successfully received as long
as it does not overlap or \say{collide} with another. Although
this model is analytically amenable and reflected the state
 of technology when networking was an emerging field, the
classical collision model does not represent the capabilities of
today's transceivers. In particular, present transceiver technologies
enable users to correctly receive multiple simultaneously
transmitted data packets. With proper design, this capability --- commonly referred to as multiple packet reception (MPR)
\cite{gvs2,gvs} --- can significantly enhance network performance.}.

Communication models allowing multiple simultaneous successful transmissions  have received great attention in the literature in recent times \cite{ chan_berg, tcs1,adua,gg, russ,tsyb}. 
The fundamental question that arises when studying  conflict resolution in the above described models is whether 
it is possible to resolve conflicts  in a number of time slots linearly decreasing with the number $d$ of messages that can be simultaneously transmitted with success.
In this paper we give a positive answer to this question for  multiple-access systems with feedback, i.e., systems  in which whenever an active station transmits to the channel, it receives a feedback that informs the station on whether its transmission has been successful.

More specifically, 
we consider a multiple-access system in which $n$ stations have access to the channel and at most $k\leq n$ stations are willing to transmit a message at the same time. We call these stations {\em active} stations. If at most $d\leq k$ active stations transmit to the channel then these stations succeed to transmit their messages, whereas if more than $d$ stations transmit then all messages are lost.   
In this latter case, we say that a {\em conflict} occurs. We assume that time is divided into time slots and that transmissions occur during these time slots. We also assume that all stations have a global clock and that active stations start transmitting at the same time slot.
 A scheduling algorithm for such a multiaccess system is a protocol that schedules the transmissions of the $n$ stations over a certain number $t$
of time slots ({\em steps}) identified by integers $1,2,\ldots, t$.
Whenever an active station transmits to the channel, it receives a feedback from the channel that informs the station on whether its transmission has been successful. As soon as an active station becomes aware that its message has been successfully transmitted,
 it becomes {\em inactive} and does not transmit in the following time slots, even though it is scheduled to transmit by the protocol. 
For the particular case $d=1$, our model corresponds to the multiple-access model with feedback considered by Koml\'os and Greenberg  in \cite{kog}.

In this paper we focus on non-adaptive scheduling algorithms, that is, algorithms that schedule all transmissions in advance so that all stations transmit according to a predetermined protocol known to them from the very beginning.  Please notice that the knowledge of the feedback cannot affect  the  schedule of transmissions but can only  signal a station to become inactive after it has successfully transmitted.

A non-adaptive scheduling algorithm is represented by a  $t\times n$
Boolean matrix where each column is associated with a distinct station and a  station $j$ is scheduled to transmit at step $i$ if and only if  entry $(i,j)$ of the matrix  is
1. In fact station $j$ really transmits at step $i$ if and only it is an active station and  is scheduled to transmit at that step. 
A  {\em conflict resolution algorithm} is a scheduling protocol that schedules transmissions in such a way that all active stations transmit with success, i.e., for each active station there is a time slot in which it is scheduled to transmit on the channel and at most $d-1$ other active stations are allowed to transmit in that time slot.  The conflict resolution protocols considered in this paper are non-adaptive. The parameter we are interested in minimizing is the number of rows of the matrix which corresponds to the  number of time slots over which the conflict resolution algorithm schedules the transmissions.  
For the case $d=1$, Koml\'os and Greenberg   \cite{kog} gave a non-adaptive protocol that  uses $O(k\log (n/k))$ time slots  to solve all conflicts among up to $k$ active stations. Later on, the authors of \cite{dgv,ko} proved the same upper bound by providing a  simple construction based on selectors \cite{dgv}. The above upper bound has been shown to be the best possible in   \cite{cohen}, and later on,
independently by the authors  of \cite{cms,csurus}. The 
lower  bound in \cite{cms,cohen,  csurus} improved on the  $\Omega(k(\log n)/(\log k))$ lower bound in \cite{gwin}, which additionally holds  for adaptive algorithms that however are not the topic of this paper.
\subsection{Our results}
In this paper we investigate the conflict resolution problem under the multiaccess model described in the previous section. To this aim, we introduce a new generalization of the codes introduced by Koml\'os and Greenberg in \cite{kog}. We prove that these new codes are equivalent  to scheduling algorithms that allow up to $k$ active stations to transmit with success in our setting, thus showing that  upper and lower bounds on the minimum length of these codes translate  into upper and lower bounds on the minimum number of time slots needed to solve conflicts. 
We present upper and lower bounds of the minimum length of these codes that differ asymptotically by a $\log (k/d)$ factor. These bounds are a consequence of the corresponding bounds for other two new combinatorial structures also introduced in this paper. The proposed construction is based on  a new  version of  $(k,m,n)$-selectors \cite{dgv} having an additional parameter $d$.
The lower bound follows from a non-existential result for  a new combinatorial structure that can be regarded as an extension of the  selective families of \cite{bbch,cggpr} and  the  $\leq $ $k$-locally thin codes of \cite{csurus}. We call these new structures $(\leq k,d,n)$-locally thin codes.

Our paper is organized as follows. In Section \ref{sec:defi} we introduce the fundamental combinatorial tools. We first introduce a generalization of  Koml\'os and Greenberg codes and prove that these new codes are equivalent to conflict resolution algorithms for our problem. Then, we introduce our generalized version of selectors and describe how to obtain a conflict resolution protocol  by concatenating generalized selectors with properly chosen parameters. We conclude Section \ref{sec:defi} by 
giving the definition of  $(\leq k,d,n)$-locally thin codes. We  show  that our generalized version of  Koml\'os and Greenberg codes is indeed  a $(\leq k,d,n)$-locally thin code, thus proving that any non-existential result for $(\leq k,d,n)$-locally thin codes imply a non existential result for the conflict resolution protocols.
In Section \ref{sec:bounds} we provide  constructions of  generalized selectors and  exploit it to obtain a construction for our  version of Koml\'os and Greenberg codes. In that section we also give  a lower bound on the minimum length of $(\leq k,d,n)$-locally thin codes. Moreover, we obtain a non-existential result for  a combinatorial structure satisfying a weaker property than that of $(\leq k,d,n)$-locally thin codes and that can be regarded as a generalization of the $k$-locally thin families of \cite{afk}.  Besides its combinatorial interest, this result implies a lower bound on the number of times slots needed to solve conflicts when the number of active stations is known to be  {\em exactly} equal to $k$. 
 Our main results are summarized by the following theorems.
 \begin{theorem}\label{thm:mainUpper}
Let $k$, $d$, and $n$ be  integers such that $1\leq d\leq  k\leq n$.  There exists a conflict resolution algorithm for a multiple-access channel with feedback that schedules the transmissions of
$n$ stations in such a way that all active stations transmit with success, provided that the number of active stations is at most $k$ and that the channel allows up to $d$ stations to transmit their messages simultaneously with success. The number of time slots $t$ used by this algorithm is 
$t=O\left(\frac kd\log \frac nk\right).$
\end{theorem}
\begin{theorem}\label{thm:mainLower}
Let $k$, $d$, and $n$ be positive integers such that $3(d+1)\leq k\leq n$.  Let ${\cal A}$ be {\em any} conflict resolution algorithm for a multiple-access channel with feedback that schedules the transmissions of
$n$ stations in such a way that all active stations transmit with success, provided that the number of active stations is at most $k$ and that the channel allows up to $d$ stations to transmit their messages simultaneously with success. The number of time slots $t$ needed by  ${\cal A}$  is 
$t=\Omega\left(\frac k{d\log(k/d)}\log (\frac n{k(d+1)}\right )$.
\end{theorem}
We remark that the asymptotic upper bound of Theorem \ref{thm:mainUpper} holds also in the case when there is  no a priori knowledge of the number $k$ of active stations. In this case, conflicts are resolved by running the conflict resolution algorithm of Theorem \ref{thm:mainUpper} iteratively (in stages), each time doubling the number of stations that are assumed to be active. In other words, at stage $i$  the conflict resolution algorithm of  Theorem \ref{thm:mainUpper} is run for a number $k_i$ of supposedly active stations equal to $2^i$. At stage $\lceil \log k\rceil$, the algorithm of Theorem \ref{thm:mainUpper} is run for a number of active stations larger than or equal to $k$ and we are guaranteed that all active stations transmit with success  within that stage.
\section{Combinatorial Structures}\label{sec:defi}
In the following, for a positive integer $m$, we denote by $[m]$ the set  $\{1,2,\ldots,m\}$.
Given a matrix $M$, we denote the set of its columns and the set of its column indices by  $M$ itself. The rows of a $t\times n$ matrix are numbered from the top to the bottom with integers from 1 to $t$.  The $n$ stations are identified by integers from 1 through $n$
and for a given subset $S\subseteq [n]$ and an $n$-column matrix $M$, we  denote by  $M[S]$ the submatrix formed by the columns of $M$ with indices in $S$.
\begin{definition}\label{def:kdn_KG}
Let $k$, $d$, and $n$ be  integers such that $1\leq d\leq  k\leq n$. A $t\times n$ Boolean matrix  is said to be a KG $(k,d,n)$-code of length $t$ if for any submatrix $M'$ of $k$ columns of $M$ there exists  a non-empty set of  row indices $\{i_1,\ldots,i_\ell\}\subseteq [t]$, with $i_1\leq i_2\leq\ldots\leq i_\ell$, such that the following property holds.
\begin{itemize}
\item[] There exists a partition $\{M'_1,\ldots,M'_\ell\}$ of the set of columns of $M'$ such that, for $j=1,\ldots,\ell$, $1\leq |M'_j|\leq d$ and
the $i_j$-th row of $M'$ has all entries at the intersection with the columns of $M'_j$ equal to 1 and those at the intersection with the columns in
$M'_{j+1},\ldots,M'_\ell$ equal to 0. 
\end{itemize}
We will denote by  $t_{KG}(k,d,n)$ the minimum length of a KG $(k,d,n)$-code.
\end{definition}
\begin{theorem}\label{thm:equivalence}
A scheduling algorithm solves all conflicts among up to $k$, $k\leq n$, active stations if and only if the corresponding Boolean matrix is a KG $(k,d,n)$-code. 
\end{theorem}
\proof
We first prove the ``if" part. Let $S$ be an arbitrary set of  $|S|\leq k\leq n$  active stations and let $M$ be a KG $(k,d,n)$-code of length $t$. We will show that $M$ provides us with a scheduling algorithm that allows all stations in $S$ to transmit with success. To this aim let us consider a $k$-column submatrix $M'$ of $M$ such that  $M[S]\subseteq M'$, i.e., $M'$ contains all columns corresponding to the stations in $S$. Moreover, let the row indices $i_1,\ldots,i_\ell$ and the submatrices $M'_1,\ldots,M'_\ell$ of $M'$ be defined as in Definition \ref{def:kdn_KG}, and for $j=1,\ldots, \ell$, let  $S_j$ be the subset of the stations of $S$ that are associated with columns in $M'_j$, i.e.,  it holds  $M[S_j]=M[S] \cap M'_j$.
By Definition \ref{def:kdn_KG}, one has that  the $i_j$-th row of $M'$ has all entries at the intersection with the columns of $M'_j$ equal to 1 and those at the intersection with the columns in
$M'_{j+1},\ldots,M'_\ell$ equal to 0. 
As a consequence, for $j=1,\ldots,\ell$, all stations in $S_j$ are scheduled to transmit in time slot $i_j$, whereas for each  $q>j$ all stations in $S_q$ are not allowed to transmit.  Moreover, by Definition \ref{def:kdn_KG}, it is $1\leq |M'_j|\leq d$. We will prove by induction that this fact implies   that all stations in $S_j$ transmit with success in time slot $i_j$. Observe that all stations in $S_1$ transmit with success in time slot $i_1$  since  $|S_1|\leq |M'_1|\leq d$ and all other active stations are in some subset $S_q$ with $q>1$. Assuming by induction that all stations in $S_p$ transmit with success in time slot $i_p$, for $1\leq p\leq j-1$,  it follows that all stations in $S_j$ transmit with success in time slot $i_j$ because   $|S_j|\leq |M'_j|\leq d$ and, by induction hypothesis,  any other station, among those that are still  active, is  contained in some subset $S_q$ with $q>j$. By Definition \ref{def:kdn_KG},  $\{M'_1,\ldots,M'_\ell\}$ is a partition  of $M'$, and consequently $\{S_1,\ldots,S_\ell\}$ is a partition of $S$, thus implying that each station in $S$ transmits with success in one of time slots $i_1,\ldots,i_\ell$.

 In order to prove the ``only if" part, let us consider  an $n\times t$ Boolean matrix $M$ corresponding to a conflict resolution algorithm  for our model and let $M'$ be an arbitrarily chosen $k$-column submatrix of
 $M$. We will show that  $M'$ satisfies the  property in Definition \ref{def:kdn_KG}.
Let us denote by $S$ the set of stations corresponding to the columns of $M'$. For  $i=1,\ldots,t$, we denote by $S_{i}$, $S_i\subset S$, the subset of active stations that transmit with success in time slot $i$ when the set of active stations is $S$.  In order for the active stations in $S_{i}$ to transmit with success, it must be  $|S_{i}|\leq d$, and any active station $r\in S_{i+1}\cup\ldots\cup S_{t}$  must not be allowed to transmit at time slot $i$ since it is still active at that time, and therefore, if allowed to transmit, it would either cause a conflict with the stations in $S_{i}$,
or it would be in fact $r\in S_{i}$, thus contradicting the assumption that 
  $r$ transmits with success after time slot $i$.
Let $S_{i_1},\ldots,S_{i_\ell}$ denote the non-empty members of $\{S_1,\ldots, S_t\}$ and let $M'_{i_1},\ldots,M'_{i_\ell}$ be the submatrices of $M$  whose columns correspond to the stations in $S_{i_1},\ldots,S_{i_\ell}$, respectively, that is, $M'_{i_j}=M[S_{i_j}]$, for $j=1,\ldots,\ell$. From the above discussion it follows that for $j=1\ldots,\ell$, $|M'_{i_j}|\leq d$ and that the $i_j$-th row of $M'$ has all entries at the intersection with the columns of $M'_{i_j}$ equal to 1 and those at the intersection with the columns in
$M'_{i_{j+1}},\ldots,M'_{i_\ell}$ equal to 0. 
 In order to show that $M'$ satisfies the property of Definition \ref{def:kdn_KG}, we need to show also  that $M'_{i_1},\ldots,M'_{i_\ell}$  form a partition of the set of columns of $M'$. To this aim, notice  that  each station $r\in S$ transmits with success exactly once, and consequently each station in $S$ is contained exactly in one members of $\{S_{i_1},\ldots,S_{i_\ell}\}$. This proves that  $\{S_{i_1},\ldots,S_{i_\ell}\}$ is a partition of $S$, and consequently $\{M'_{i_1},\ldots,M'_{i_\ell}\}$ is a partition of $M'$, thus concluding the proof that $M'$ satisfies the property of Definition \ref{def:kdn_KG}.  
\qed
The following definition introduces a new combinatorial structure that will be employed as a building block  to construct  KG $(k,d,n)$-codes. This new structure generalizes the notion of $(k,m,n)$-selectors introduced in \cite{dgv} and corresponds to this notion for  $d=1$.  
\begin{definition}\label{def:kdn_sel}
Let $k$, $m$, $d$, and $n$ be  integers such that $1\leq d\leq m\leq  k\leq n$. A $t\times n$ Boolean matrix is  said to be a $(k,m,d,n)$-selector
if any $k$-column submatrix $M'$ of $M$ contains a set $R$ of rows such that each row in $R$ has Hamming weight comprised between 1 and $d$, and the Boolean sum of all rows of $R$ has Hamming weight at least $m$. The number of rows $t$ of the $(k,m,d,n)$-selector is the size of the selector. The minimum size   of $(k,m,d,n)$-selectors is denoted by  $t_{sel}(k,m,d,n)$.
\end{definition}
A $(k,m,d,n)$-selector defines a scheduling algorithm for our multiaccess model that, in the presence of 
up to $k$ active stations, allows all but at most $k-m$ of these stations to transmit with success. Indeed, all active stations that are  scheduled to transmit in the  time slots corresponding to the rows in $R$, transmit with success, since for each of those time slots there are at most $d$ stations scheduled to transmit in that time slot. Notice that
an active station might be scheduled to transmit in more than one of those time slots but it will become inactive as soon as it transmits with success. Let $p\leq k$ be the total number of active stations. Since the Boolean sum of the rows in $R$ has Hamming weight at least $m$,   then at least $m-(k-p)$  1-entries in that  Boolean sum  are associated with active stations, and consequently, at least $m-(k-p)$  active stations transmit with success and at most  $k-m$ active stations do not succeed to transmit their messages. 

In the following we will show how to use $(k,m,d,n)$-selectors to obtain a  KG $(k,d,n)$-code. The idea of this construction is similar to the  one employed in \cite{dgv,ko} to obtain a  KG $(k,1,n)$-code by using  $(k,m,n)$-selectors as building blocks.
From now on,
 unless specified differently, ``$\log$" will denote the logarithm in base 2.
For the moment, let us assume for the sake of the simplicity that $k$ and $d$ be  powers of 2.  Our construction works as follows. We concatenate the rows of $(2^{v+1},2^{v},d,n)$-selectors, for $v=\log d,\ldots, \log k-1$,
with the rows of the $(k,k/2,d,n)$-selector  being placed  at the top and those of the $(2d,d,d,n)$-selector being placed at the bottom. Then we  add an all-1 row at the bottom of the matrix. Let $M$ be the resulting matrix.  Notice that the protocol defined by $M$ consists in running the protocols defined by the $(2^{v+1},2^{v},d,n)$-selectors, starting from the protocol associated with the $(k,k/2,d,n)$-selector through the one associated with the
$(2d,d,d,n)$-selector.  In the last time slot, corresponding to the bottommost row of $M$, the protocol schedules all stations to transmit.  
 We will show that $M$  defines a scheduling algorithm the allows up to $k$ active stations to transmit with success, which, by Theorem \ref{thm:equivalence}, is equivalent to showing that $M$ is a KG $(k,d,n)$-code. 
Let $S$ be an arbitrary set of up to $k$ active stations. 
We observed that a $(k,m,d,n)$-selector provides a scheduling algorithm that schedules the transmissions so that  at most $k-m$ active stations do not succed to transmit their messages.
Therefore, after running the scheduling protocol associated with the $(k,k/2,d,n)$-selector, the algorithm is left with at most $k/2$ active stations. Then the algorithm runs the protocol associated with the $(k/2,k/4,d,n)$-selector. This protocol allows all but at most $k/4$ of the remaining active stations to transmit with success. Let $t_v$ denote the number of rows of the $(2^{v+1},2^{v},d,n)$-selector, for $v=\log d,\ldots, \log k-1$.   For an arbitrary $v$, we have that after $t_{\log k-1}+\ldots+t_{v}$ time slots there are at most $2^v$ stations that are still active. Therefore, after running the protocol associated with the $(2d,d,d,n)$-selector, there are most $d$ active stations and no conflict can occur in the last time slot. In the last time slot all stations are scheduled to transmit, and consequently,  all remaining active stations transmit with success in that time slot.
For arbitrary values of $k$ and $d$ (not necessarily  powers of 2), we replace in the above construction $\log k$ and $\log d$ by $\lceil \log k\rceil$ and $\lfloor \log d\rfloor$, respectively. 
The above construction implies the following upper bound on the minimum length $t_{KG}(k,d,n)$ of a $KG$ $(k,d,n)$-code 
\begin{equation}\label{eq:upper_KG1}
t_{KG}(k,d,n)= O\Big(\sum_{i=\lfloor \log d\rfloor}^{\lceil \log k\rceil-1}t_{sel}(2^{i+1},2^i,d,n)\Big).
\end{equation}
Below, we define a novel combinatorial structure that is strictly related to our problem in that non-existential results for this structure translate into non-existential results for KG $(k,d,n)$-codes.
\begin {definition}\label{def:loc_thin}
Let $k$, $d$, and $n$ be  integers such that $1\leq d\leq k\leq n$. A     $t\times n$  Boolean matrix $M$ is said to be a $(\leq k,d,n)$-locally thin code of length $t$ if the submatrix formed by any subset of $s$, $d\leq s\leq k$, columns of $M$ contains a row with a number of 1's comprised between 1 and $d$.
We will denote by  $t_{LT}(\leq k,d,n)$ the minimum length of a $(\leq k,d,n)$-locally thin code.
\end{definition}
 Let $M$ be a $(\leq k,d,n)$-locally thin code and let $\F$ be the family of the sets whose characteristic vectors are the columns of $M$.  The familiy $\F$ has the property  that for any subfamily $\F'\subseteq \F$ with $d \leq |\F'|\leq k$, there exists an element $x\in[t]$ such that $1\leq |\{F\in\F':\, x\in F\}|\leq d$.
For $d=1$, these families correspond to the selective families of \cite{bbch,cggpr} and to the $\leq$ $k$-locally thin families  of \cite{csurus}.
The authors of  \cite{cms,cohen, csurus} proved an $\Omega(k\log (n/k))$ lower bound on the minimum size  of the ground set of  $\leq$ $k$-locally thin families which is tight with the upper bound on the length of KG $(k,1,n)$-code \cite{kog}.
The following theorem establishes a relation between $(\leq k,d,n)$-locally thin codes and KG $(k,d,n)$-codes.
\begin{theorem}\label{thm:forLower}
Let $k$, $d$, and $n$ be  integers such that $1\leq d\leq k\leq n$. Any $KG$ $(k,d,n)$-code  is a  $(\leq k,d,n)$-locally thin code.
\end{theorem}
\proof
Let $M$  be a $KG$ $(k,d,n)$-code and suppose by  contradiction that $M$ is not a $(\leq k,d,n)$-locally thin code. This implies that there exists a
 subset of $s$, $d\leq s\leq k$, columns of $M$ such that the submatrix $M_s$ formed by these $s$ columns contains no row with a number of 1's comprised between 1 and $d$. Let $M'$ be a $k$-column submatrix of $M$ such that $M'\supseteq M_s$. Since $M$ is a KG $(k,d,n)$-code,    Definition \ref{def:kdn_KG} implies that  there exists  a non-empty set of  row indices $\{i_1,\ldots,i_\ell\}\subseteq [t]$, with $i_1\leq i_2\leq\ldots\leq i_\ell$, such that the following property holds
 \begin{itemize}
\item[] There exists a partition $\{M'_1,\ldots,M'_\ell\}$ of the set of columns of $M'$ such that, for $j=1,\ldots,\ell$,  $1\leq |M'_j|\leq d$ and
the $i_j$-th row of $M'$ has all entries at the intersection with the columns of $M'_j$ equal to 1 and those at the intersection with the columns in
$M'_{j+1},\ldots,M'_\ell$ equal to 0. 
\end{itemize}
Let $M'_{f_1},\ldots,M'_{f_b}\in \{M'_1,\ldots,M'_\ell\}$, with $f_1\leq f_2\ldots\leq f_b$, be the members of the partition having non-empty intersection with $M_s$, i.e., $M_s\cap (M'_{f_1}\cup\ldots \cup M'_{f_b})=M_s$.  
By our assumption that $M_s$ does not contain any row with Hamming weight comprised between 1 and $d$, it follows that each row of $M_s$ has either Hamming weight 0 or Hamming weight larger than $d$. In the former case, the row has a 0   in correspondence of at least one column in each of $M'_{f_1},\ldots,M'_{f_b}$, whereas in the latter case  the row has  entries equal to 1 in correspondence of  columns belonging to at least two of $M'_{f_1},\ldots ,M'_{f_b}$, since these submatrices contain  at most $d$ columns. Let us consider the row of $M_s$ with index $i_{f_1}$. By  Definition \ref{def:kdn_KG}, one has that this row  has the entries at the intersection  with the columns in $M'_{f_1}$ equal to 1 and those at the intersection with the columns in
$M'_{f_2}\cup\ldots \cup M'_{f_b}$ equal to 0. However, from what we have just observed,  the $i_{f_1}$-th row of $M_s$ has either a 0  in correspondence of at least one column in each of $M'_{f_1},\ldots,M'_{f_b}$, or has  entries equal to 1 in correspondence of columns belonging to at least two of $M'_{f_1},\ldots,M'_{f_b}$. In the former case,  the $i_{f_1}$-th row of $M_s$ has an entry equal to 0 also at the intersection with some column in $M'_{f_1}$, whereas in the latter case the $i_{f_1}$-th row has a 1-entry in correspondence of some column in at least one of $M'_{f_2},\ldots, M'_{f_b}$. In both cases, the $i_{f_1}$-the row of $M_s$ does not satisfy the property of Definition \ref{def:kdn_KG},  thus contradicting the fact that $M$ is a 
$KG$ $(k,d,n)$-code.
\qed
%
\section{Upper and lower bounds}\label{sec:bounds}
The following theorem provides an upper bound on the minimum size   of $(k,m,d,n)$-selectors for $k> 2(m-1)$. 
\begin{theorem}\label{thm:upper1}
Let $k$, $m$, $d$, and $n$ be positive integers such that $1\leq d\leq m$ and $2(m-1)<  k\leq n$. The minimum size  $t_{sel}(k,m,d,n)$ of  a $(k,m,d,n)$-selector
is $$t_{sel}(k,m,d,n)\leq \begin{cases}  16(k\ln\left({n\over k}\right)+(k-m+1)\ln\left({k\over k-m+1}\right)+2k-m+1)& \mbox{ if }
1\leq d\leq 2\cr
{k\ln\left({n\over k}\right)+(k-m+1)\ln\left({k\over k-m+1}\right)+2k-m+1\over {d(k-m+1)\over 4k}-\ln (4/3)}&\mbox{ otherwise, }\end{cases}$$
where $e $ denotes the Neper's constant $e = 2,71828\ldots $. 
\end{theorem}
\proof
We will prove the existence of a $(k,m,d,n)$-selector with size  smaller than or equal to the stated upper bound. The proof is by the probabilistic method. 
Let $M$ be a $t\times n$ random binary matrix $M$ where each entry is 1 with probability $p$ and 0 with probability
$1-p$. We want to estimate the probability that  $M$ is not a $(k,m,d,n)$-selector.  To this aim we compute an upper bound on the probability $P$ that  a given $k$-column submatrix $M'$ of $M$ does not satisfy the property of Definition \ref{def:kdn_sel}.
In the following we will say that a row is $w$-good if its Hamming weight is comprised between 1 and $d$.
The probability $P$ is the probability that the  submatrix $M'$ contains no subset $R$ of $w$-good rows  such that the Boolean sum of the rows in $R$ has Hamming weight larger than or equal to $m$. To this aim, we notice that this event holds if and only if
there exists a set $A$ of $k-m+1$ column indices  such that all $w$-good rows of $M'$ have all zeros at the intersection with the columns with indices in $A$.
For a fixed subset $A$ of $k-m+1$ indices of columns of $M'$, we denote by $E_A$ the event that each $w$-good row of $M'$ has zeros at the intersection with the columns with indices in $A$. Hence, we have that 
\begin{equation}\label{eq:P}
P=Pr\Big\{\bigcup_{\small \begin{array}{c} \mbox{$A\subseteq M'$:}\cr
|A|=k-m+1  \end{array}} E_A\Big\}\leq \sum_{\small \begin{array}{c} \mbox{$A\subseteq M'$:}\cr
|A|=k-m+1  \end{array}}  Pr\{E_A\},
\end{equation}
and 
\begin{equation}\label{eq:badP}
Pr\{\mbox{$M$ is not a $(k,m,d,n)$-selector}\}\leq {n\choose k}\sum_{\small \begin{array}{c} \mbox{$A\subseteq M'$:}\cr
|A|=k-m+1  \end{array}}  Pr\{E_A\}.
\end{equation}
For a fixed $A$, event $E_A$ holds if and only if, for any row index $i=1,\ldots,t$,  one has either that the $i$-th row of $M'$ is not $w$-good or that the $i$-th row of $M'$ is $w$-good and has all zeroes at the intersection with  the   columns with indices in   $A$. Therefore, one has that
\begin{eqnarray}\label{eqE_A}
Pr\{E_A\}&=& Pr\Big\{\bigcap_{i=1}^t  \big\{\{\mbox{the $i$-th row of $M'$ is not $w$-good}\}\cr
&&\!\!\!\!\!\!\!\!\!\!\!\!\!\!\cup \{\mbox{the $i$-th row of $M'$  is $w$-good and $M'(i,j)=0$ for all $j\in A$} \}\big\}
\Big\}\leq (P_1+P_2)^t, 
\end{eqnarray}
where \begin{equation}\label{eq:defP1}P_1=Pr\{\mbox{the $i$-th row of $M'$ is not $w$-good}\big\}\end{equation} and
\begin{equation}\label{eq:defP2}P_2= \{\mbox{the $i$-th row of $M'$  is $w$-good and $M'(i,j)=0$ for all $j\in A$}\},\end{equation}
 for a fixed $i\in [t]$. Notice that  $P_1$ and $P_2$ do not depend on $i$.
By (\ref{eq:badP}) we have that
\begin{equation}
Pr\{\mbox{$M$ is not a $(k,m,d,n)$-selector}\}\leq {n\choose k}{k\choose k-m+1}(P_1+P_2)^t.
\end{equation}
The above probability is strictly smaller than 1 for 
\begin{equation}\label{eq:boundt}
t>{\ln\left({n\choose k}{k\choose k-m+1}\right)\over -\ln(P_1+P_2)}.
\end{equation}
Therefore, there exists a  $(k,m,d,n)$-selector of size $t$, for any $t$ satisfying the above inequality.

\medskip
The proof of  the following claim is given in the Appendix.

\medskip\noindent
{\bf Claim 1.} Let $k$, $m$, $d$, and $n$ be positive integers such that $1\leq d\leq m$ and $2(m-1)<  k\leq n$. It is possible to choose  $p\in(0,1)$ in such a way that it holds 
$$ -\ln(P_1+P_2)\geq 
\begin{cases} \frac 1 {16} & \mbox{ if }
d\in\{1,2\}\cr
(k-m+1)\left(\frac d{4k}\right)-\ln(\frac 43) &\mbox{ if $d\geq 3$.}\end{cases}$$

\medskip
In order for a $(k,m,d,n)$-selector of size $t$ to exist it is sufficient that 
$t$ satisfies inequality (\ref{eq:boundt}).  Claim 1 implies that  the righthand side of (\ref{eq:boundt}) is strictly smaller than
$16( \ln{n\choose k}+\ln{k\choose k-m+1})$, if $1\leq d\leq 2$, and it is strictly smaller than
 ${\ln{n\choose k}+\ln{k\choose k-m+1}\over {d(k-m+1)\over 4k}-\ln (4/3)}$, if $d\geq 3$. 
The upper bounds in the statement of the theorem are a consequence of these two upper bounds and of  the following well known upper bound on the binomial coefficient: ${z\choose y}\leq \left( {ez\over y}\right)^y$.
\qed
Theorem \ref{thm:upper1} implies that bound  (\ref{eq:upper_KG1}) on $t_{KG}(k,d,n)$ in Section \ref{sec:defi} is
$O\Big(\sum_{i=\lfloor \log d\rfloor}^{ \lceil \log k\rceil -1} \frac{2^{i+1}}d\log (\frac n{2^{i+1}})\Big)=O\left(\frac k  d\log \frac nk\right).$ Therefore, the following theorem holds.

\begin{theorem}\label{thm:upperLT}
Let $k$, $d$, and $n$ be positive integers such that $d\leq  k\leq n$.  The minimum length $t_{KG}(k,d,n)$ of a KG $(k,d,n)$-code is
$t_{KG}(k,d,n)=O((k/d)\log (n/k)).$
\end{theorem}
Theorem \ref{thm:mainUpper} follows from Theorems \ref{thm:equivalence} and \ref{thm:upperLT}.
In virtue of Theorem \ref{thm:forLower}, we have that Theorem \ref{thm:upperLT} implies an existential results for $(\leq k,d,n)$-locally thin code. For $d=1$, this existential result attains the same asymptotic upper  bound as the one in \cite{cms}.
The following theorem states a lower bound on the minimum length of $(\leq k,d,n)$-locally thin codes.
\begin{theorem}\label{thm:lower<=}
Let $k$, $d$, and $n$ be positive integers such that $3(d+1)\leq k\leq n$. The minimum length  $t_{LT}(\leq k,d,n)$ of  a $(\leq k,d,n)$-locally thin code is
$t_{LT}(\leq k,d,n) >  {\big\lfloor {k\over d+1}\big\rfloor  \over \log\left(e \big\lfloor {k\over d+1}\big\rfloor\right)}\log \left(\frac{n}{k(d+1)}\right).$
\end{theorem}
\proof 
Let us  write $k$ as $k=(d+1)\lfloor\frac k{d+1}\rfloor+q$, with $0\leq q\leq d$ and let $u=\lfloor\frac k{d+1}\rfloor$. We denote by $\alpha$ a positive rational number $\alpha=\frac a b$ satisfying the following inequalities
\begin{equation}\label{eq:alpha1}
 \frac1u\leq \alpha<\frac 1 2.
\end{equation}
Let us denote by $n_{LT}(\leq k,d,t)$ the maximum value of $n$ for which there exists a $(\leq k,d,n)$-locally thin code of length $t$.
We will prove  that 
$n_{LT}(\leq k,d,t)< k(d+1)\cdot 2^{h(\alpha)t}$. 
First we will  show that for any $\alpha<\frac 1 2$ it holds
\begin{equation}\label{eq:alpha2}
\frac{\alpha} e \leq 2^{-\frac {h(\alpha)}{\alpha}}<\frac \alpha 2,
\end{equation}
where $h(\alpha)$ denotes the binary entropy of $\alpha$. Notice that, since we can  choose $\alpha = \frac1u$,  the upper bound on $n_{LT}(\leq k,d,t)$, along with the lefthand side of (\ref{eq:alpha2}), implies the lower bound on  $t_{LT}(\leq k,d,n)$ in the statement of the theorem.
Let us prove inequalities (\ref{eq:alpha2}).
By the definition of binary entropy, one has that
\begin{eqnarray}
h\left(\frac a b\right)&=&\frac a b \log \frac b{a}+\left({b-a\over b}\right)\log\left({b\over b-a}\right)
=\frac a b \log \frac b{a}+{1\over b}\cdot\log\left(1+{{a}\over b-a}\right)^{b-a} \label{eq:en}.\end{eqnarray}
Since $\left(1+{a\over b-a}\right)^{b-a}$ increases with $b$, one has that $2^{a}<\left(1+{a\over b-a}\right)^{b-a}\leq e^{a}$,
where the left inequality follows from the righthand side (\ref{eq:alpha1}) that implies $b>2a$.
Therefore, by (\ref{eq:en}), it holds
\begin{equation}\label{eq:sand1}
\frac a b \log\left(\frac {2b}{a}\right)<  h\left(\frac a b\right)\leq \frac a b \log\left(\frac {eb}{a}\right).
\end{equation}
By replacing $\frac a b$ with $\alpha$, inequalities (\ref{eq:sand1}) can be rewritten as 
$\alpha \log\left(\frac {2}{\alpha}\right)<  h(\alpha)\leq  \alpha \log\left(\frac {e}{\alpha}\right)$,
from which we have that inequalities (\ref{eq:alpha2}) hold.

Now we prove that  $n_{LT}(\leq k,d,t)< k(d+1)\cdot 2^{h(\alpha)t}$. The proof is by induction on $t$. 
For $t=1$, any $t\times n$ Boolean matrix $M$ has a single row that either contains at least $\frac n2$ entries equal to 0 or at least $\frac n 2$ entries equal to 1. Consequently, if we assume by contradiction that
$|M|=n\geq  k(d+1)\cdot 2^{h(\alpha)t}\geq k(d+1)$ then the single row of $M$ would either contain at least $k(d+1)/2$ occurrences of 0 or at least $k(d+1)/2$ occurrences of 1. This implies that there exist $k(d+1)/2\geq k$ entries that  are either all equal to 0 or all equal to 1 thus contradicting the hypothesis that $M$ is a $(\leq k,d,n)$-locally thin code.

Let us consider $t>1$ and let us assume by induction hypothesis that $n_{LT}(\leq k,d,t-1)< k(d+1)\cdot 2^{h(\alpha)(t-1)}$.
Let $M$ be a $t\times n$ be a  $(\leq k,d,n)$-locally thin code  of length $t$  and let us assume by contradiction that $n\geq k(d+1)\cdot 2^{h(\alpha)t}$.
We consider the following two cases.
\begin{itemize}
\item {Case 1.} There exists an integer $i$ in $[t]$ such that there are at least $2^{-h(\alpha)}n$ columns of $M$ with the $i$-th entry equal to 0.
If we remove the $i$-th entry from each of these columns, we have that the resulting  columns form a matrix $\Mt$    that is a  $(\leq k,d,n)$-locally thin code of length $t-1$. Since we are assuming that  $n\geq k(d+1)\cdot 2^{h(\alpha)t}$, it holds $|\Mt|\geq 2^{-h(\alpha)}k(d+1)2^{h(\alpha)t}= k(d+1)\cdot 2^{h(\alpha)(t-1)}$.
By induction hypothesis,  $\Mt$ cannot be a $(\leq k,d,n)$-locally thin code of length $t-1$, thus contradicting the fact that $M$ is $(\leq k,d,n)$-locally thin code.
\item{Case 2.} For each element $i\in  [t]$, there are less than $2^{-h(\alpha)}n$ columns of $M$ with the $i$-th entry equal to 0.
This implies that for a fixed $i$ and for  $u$ randomly chosen columns $\bc_1,\ldots,\bc_{u}$
of $M$, the probability that $\bc_1,\ldots,\bc_{u}$ all have the $i$-th entry equal to 0  is less than  $2^{-{u}h(\alpha)}$. 
By the lefthand side of (\ref{eq:alpha1}) this probability is at most $2^{-\frac {h(\alpha)}{\alpha}}$, which by the righthand side of (\ref{eq:alpha2}) is less than  $\frac \alpha 2$.
Therefore, the expected number of 0-entries in the Boolean sum $\bigvee_{j=1}^u\bc_j$ is  less than   $\frac {t\alpha} 2$.
Let $X$ denote the number of 0-entries in the Boolean sum of $u$ randomly chosen columns. We have shown that
$E[X]< \frac {t\alpha} 2.$
Markov's inequality implies that, for any non-negative random variable $Y$ and for any $b>0$, it holds 
$Pr\{Y\geq b\}\leq \frac{E[Y]}b.$
By our upper bound on $E[X]$ and by Markov's inequality, one has 
$\Pr\{\mbox{  $\bigvee_{j=1}^{u}\bc_j$ has at least  $t\alpha$ 0-entries}\}< \frac{t\alpha} 2\cdot \frac 1{ t\alpha}=\frac 12$.
It follows that $Pr\{\mbox{$\bigvee_{j=1}^u \bc_j$ has Hamming weight larger than $t-t\alpha$}\}> \frac 12$.
Let $m=2(d+1)\lceil 2^{h(\alpha)t}\rceil$ and let $\B_1,\ldots,\B_m$ be $m$ randomly chosen subsets of $u$ columns of $M$ such that $\B_j\cap \B_\ell=\emptyset$,
for $j\neq \ell$. Such subsets $\B_1,\ldots,\B_m$ can be  generated by  randomly permuting the columns of $M$, and then picking a set of $m\cdot u$ consecutive columns in the resulting matrix.  In order to obtain $\B_1,\ldots,\B_m$, this set of columns is partitioned into $m$ disjoint subsets each consisting of $u$ consecutive columns. We have shown that $\bigvee_{\bc\in \B_\ell}\bc$ has Hamming weight larger than  $t-t\alpha$ with probability larger than $\frac 12$, and consequently,
the expected number of subfamilies $\B_j$'s among $\B_1,\ldots,\B_m$ such that $\bigvee_{F\in \B_j}F$ has Hamming weight larger than or equal to $t-t\alpha$ is at least $\frac m2$.
By linearity of expectation, there is a random choice of  $\B_1,\ldots,\B_m$ such that there are at least  $f\geq \frac m2$ subfamilies $\B'_1,\ldots, \B'_{f}$ among  $\B_1,\ldots,\B_m$ for which one has that $\bigvee_{\bc\in \B'_\ell}\bc$, for $\ell=1,\ldots,f$, has Hamming weight larger than or equal to $t-t\alpha$.
However, one has that the number of pairwise distinct binary vector of length $t$ with Hamming weight larger than or equal to $t-t\alpha$ is
\begin{equation}\label{eq:b_1}
\sum_{s=t-t\alpha}^t{t\choose s}=\sum_{s=0}^{t\alpha}{t\choose s}\leq 2^{th(\alpha)},
\end{equation}
where the last inequality follows from the  well known inequality 
$\sum_{i=0}^{b}{g\choose i}\leq 2^{gh(b/g)}$,
 holding for  $b/g  \leq 1/2$, \cite{fg} . 
 Since it is $m=2(d+1)\lceil 2^{h(\alpha)t}\rceil$, then there are at most ${m\over 2(d+1)}$ pairwise distinct vectors of Hamming weight larger than or equal to
 $t- t\alpha$. We have shown that there exist $f\geq \frac m2$ subfamilies $\B'_1,\ldots, \B'_{f}$  such that $\bigvee_{\bc\in \B'_\ell}\bc$, 
 for $\ell=1,\ldots,f$, has Hamming weight larger than or equal to $t-t\alpha$. As a consequence, for at least a binary vector $\bc_v$, there are $d+1$ sets $\B'_{j_1}\ldots,\B'_{j_{d+1}}\subseteq\{\B'_1,\ldots,\B'_f\}$  such that   $ \bigvee_{\bc\in \B'_{j_q}}\bc=\bc_v,$ for $q=1,\ldots,d+1$. In other words, $\bc_v$ occurs at least $d+1$ times among the Boolean sums $\bigvee_{\bc\in \B'_1}\bc,\ldots,\bigvee_{\bc\in \B'_f}\bc$. Therefore, the submatrix formed by the $(d+1)u=(d+1) \lfloor\frac k {d+1}\rfloor\leq k$ columns of $\B'_{j_1}\ldots,\B'_{j_{d+1}}$ is such that each row is either an all-zero row or has at least $d+1$ entries equal to 1, thus contradicting the assumption the $M$ is   a $(\leq k,d,n)$-locally thin code.
\end{itemize}
 \qed
  Theorem 
 \ref{thm:mainLower} is an immediate consequence of Theorems \ref{thm:forLower} and \ref{thm:lower<=}.
 The technique used to prove the lower bound of Theorem \ref{thm:lower<=} allows also to obtain a lower bound on the length of codes satisfying a weaker property than that of $(\leq k,d,n)$-locally thin codes. We refer to these codes as $(k,d,n)$-{\em locally thin codes}. A  $t\times n$ Boolean matrix $M$ is a $(k,d,n)$-locally thin code of length $t$ if and only if any submatrix formed by {\em exactly} $k$ columns  of $M$ contains at least a row whose Hamming weight is comprised between 1 and $d$. If we interpret the columns of such a code as the characteristic vectors of $n$ sets on the ground set $[t]$, then these sets have the property that  for any $k$ of them there exists an $i\in[t]$ that is contained in at at least one of these $k$ sets and in no more than $d$ of them. For $d=1$, these families correspond to the $k$-locally thin code of  \cite{afk}.
\begin{theorem}\label{thm:lowerAFK}
Let $k$, $d$, and $n$ be positive integers such that $ 4(d+1)\leq  k\leq n$. The minimum length  $t_{LT}(k,d,n)$ of  a $(k,d,n)$-locally thin code is
$t_{LT}(k,d,n)>  {\left(\big\lfloor {k\over d+1}\big\rfloor-1\right) \over \log\left(e \left( \big\lfloor {k\over d+1}\big\rfloor-1\right) \right)} \log \left(\frac{n}{k(d+1)}\right).$
\end{theorem}
\proof
The proof is similar to the one of Theorem \ref{thm:lower<=} with the difference that here we 
write $k$ as $k=(d+1)\left(\lfloor\frac k{d+1}\rfloor-1 \right) +d+1+q$, with $0\leq q\leq d$, and set $u=\lfloor\frac k{d+1}\rfloor-1$, and in the proof for Case 2 we need to prove the existence of a submatrix of {\em exactly } $k$ columns that does not contain any row of Hamming weight comprised between 1 and $d$. To this aim, let us consider the set of columns $\B'_{j_1},\ldots,\B'_{j_{d+1}}$ whose existence has been proved in the proof of Theorem \ref{thm:lower<=}. The subsets $\B'_{j_1},\ldots,\B'_{j_{d+1}}$ are such that the Boolean sums $\bigvee_{\bc\in \B'_{j_1}}\bc,\ldots,\bigvee_{\bc\in \B'_{j_{d+1}}}\bc$ have Hamming weight at least $t-t\alpha$, and the submatrix formed by the columns in $\B'_{j_1}\cup\ldots\cup\B'_{j_{d+1}}$ contains no row of Hamming weight comprised between 1 and $d$. The number of columns in this submatrix is $(d+1)u=(d+1)\left(\lfloor\frac k {d+1}\rfloor-1\right)\leq k$. We will show that it is possible to add columns to this submatrix so as to obitan a submatrix with exactly $k$-columns  and with no row of Hamming weight comprised between 1 and $d$.
To this aim, let us consider the columns of $M$ that  do not belong to any of $\B'_{j_1},\ldots,\B'_{j_{d+1}}$. Let us denote by $i_1,\ldots, i_z$ the indices of the 0-zero entries in the Boolean sums $\bigvee_{\bc\in \B'_{j_1}}\bc,\ldots,\bigvee_{\bc\in \B'_{j_{d+1}}}\bc$. By definition of  $\B'_{j_1}\ldots,\B'_{j_{d+1}}$, one has $z\leq t\alpha$. We will prove that there are at least $d+1+q$ columns whose restrictions to the entries with indices $i_1,\ldots, i_z$ are identical. This implies that the  $k\times t$ submatrix formed by  $d+1+q$ of these columns and the columns in  $\B'_{j_1}\ldots,\B'_{j_{d+1}}$ is such that each row is either an all-zero row or has at least $d+1$ entries equal to 1, thus contradicting the fact that $M$ is a $(\leq k,d,n)$-locally thin code.
In order to prove that  there are at least $d+1+q$ columns whose restrictions to the entries with indices $i_1,\ldots, i_z$ are identical, we observe that there are at most $2^z\leq 2^{t\alpha}$ columns whose restrictions to indices  $i_1,\ldots, i_z$ are pairwise distinct. The number of columns of $M$ that  do not belong to any of $\B'_{j_1}\ldots,\B'_{j_{d+1}}$ is $n-(d+1)\left(\lfloor\frac k {d+1}\rfloor-1\right)$ that,
 by the contradiction assumption, is at least $k(d+1)\cdot 2^{h(\alpha)t}-(d+1)\left(\lfloor\frac k {d+1}\rfloor-1\right)$ and by the righthand side of (\ref{eq:alpha2}) is larger than  
 $k(d+1)\left( \frac 2 \alpha\right)^{t\alpha}-(d+1)\left(\lfloor\frac k {d+1}\rfloor-1\right)$
 Since $k(d+1)\cdot \left( \frac 2 \alpha\right)^{t\alpha}-(d+1)\left(\lfloor\frac k {d+1}\rfloor-1\right)>2d\cdot2^{t\alpha}$, it follows that there are at least $2d+1\geq d+1+q$ columns of $M$ not in $\B'_{j_1}\ldots,\B'_{j_{d+1}}$ whose restrictions to indices $i_1\ldots,i_z$ are identical.
\qed
For $k$ even, the authors of \cite{afk} proved an $\Omega(k\log n)$ lower bound on the minimum size of the ground set of $k$-locally thin families, whereas,  for arbitrary values of $k$, they gave an  $\Omega\left({k\over log k}\log n\right)$  lower bound. For $d=1$ the  bound of Theorem \ref{thm:lowerAFK} is asymptotically the same as the  bound given in \cite{afk} for arbitrary values of $k$. 
Notice that Theorem \ref{thm:lowerAFK} gives a lower bound on  the minimum number of time slots needed to solve all conflicts when the number of active stations is  {\em exactly } $k$.
\section{Conclusions} 
We have presented upper and lower bound on the minimum number of time slots needed to solve conflicts among up to $k$ active stations in a multiple-access system with feedback where at most $d$ stations can transmit simultaneously with success over the channel. Our bounds differ  asymptotically by a $\log (k/d)$ factor. An interesting  open problem is to close this gap 
 by improving on the lower bound on the minimum length of KG $(\leq k,d,n)$-codes.

\section*{Appendix} \subsection{Proof of Claim 1}
By (\ref{eq:defP1}), we have that
\begin{eqnarray}\label{eq:P_1}
P_1=Pr\{\sum_{j\in M'} M'(i,j)=0\}+Pr\{\sum_{j\in M'}  M'(i,j)\geq d+1\},
\end{eqnarray}
whereas by (\ref{eq:defP2}), we have that
\begin{eqnarray}\label{eq:P_2}
P_2&=&Pr\{\mbox{the $i$-th row of $M'$  is $w$-good $|$ $M'(i,j)=0$ for all $j\in A$} \}
\!\cdot\!
Pr\{\mbox{$M'(i,j)=0$ for all $j\in A$} \}\cr\cr
&=&Pr\Big\{1\leq \sum_{j\in M'\setminus A} M'(i,j) \leq  d\Big\}\cdot
Pr\{\mbox{$M'(i,j)=0$ for all $j\in A$} \}\cr
&=&\Big(\sum_{i=1}^d{m-1\choose i} p^i(1-p)^{m-1-i}\Big) (1-p)^{k-m+1}
=(1-p)^{k}\sum_{i=1}^d{m-1\choose i} \left({p\over 1-p}\right)^{i}.\end{eqnarray}

\medskip
\noindent
Let us consider the case $d\in\{1,2\}$.
Equality (\ref{eq:P_1})  implies
\begin{eqnarray}
P_1&=&Pr\{\sum_{j\in M'} M'(i,j)=0\}+Pr\{\sum_{j\in M'}  M'(i,j)\geq d+1\}\cr
&=&1-Pr\{1\leq \sum_{j\in M'}  M'(i,j)\leq d\} \label{app_2}=1-\sum_{i=1}^d{k\choose i} p^i (1-p)^{k-i}.
\end{eqnarray}
Equalities (\ref{eq:P_2}) and (\ref{app_2})    imply
\begin{eqnarray}
P_1+P_2&=& 1- \sum_{i=1}^d{k\choose i} p^i (1-p)^{k-i} + (1-p)^{k}\sum_{i=1}^d{m-1\choose i} \left({p\over 1-p}\right)^{i}\cr
&=&1-(1-p)^{k}\sum_{i=1}^d\left({k\choose i}- {m-1\choose i} \right)\left({p\over 1-p}\right)^{i}. \label{app_3}
 \end{eqnarray}
 Expression $(1-p)^{k}\sum_{i=1}^d\big({k\choose i}- {m-1\choose i} \big)\left({p\over 1-p}\right)^{i}$ in (\ref{app_3})
 is larger than or equal to 
 $$(1-p)^{k}(k-m+1) \left({p\over 1-p}\right). $$
 By setting $p={d\over 2k}$ in the above expression we get
 \begin{equation}\label{app_4}(1-p)^{k}(k-m+1) \left({p\over 1-p}\right)= \left(1-\frac d{2k}\right)^k (k-m+1)\left({d\over 2k-d}\right).\end{equation}
 Since $(1-\frac d{2k})^k$ increases with $k$ and $k$ is larger than or equal to $d$ it is $ (1-\frac d{2k})^k \geq \left(\frac12\right)^d$. 
 The hypothesis that $m-1<\frac k2$ implies that $k-m+1>\frac k 2$.
 Moreover, ${d\over 2k-d}$ increases with $d$, and consequently, ${d\over 2k-d}\geq {1\over 2k-1}$. The above three inequalities imply that the  righthand side of
 (\ref{app_4}) is larger than
 $$ \left(\frac12\right)^d\left(\frac k2\right)\left({1\over 2k-1}\right)\geq \left(\frac12\right)^d\frac 14.$$
 Therefore, we have that $(1-p)^{k}\sum_{i=1}^d\big({k\choose i}- {m-1\choose i} \big)\left({p\over 1-p}\right)^{i} > \left(\frac12\right)^d\frac 14 $, and by (\ref{app_3}), we get
 \begin{equation}\label{app_5}P_1+P_2< 1-  \left(\frac12\right)^d\frac 14.\end{equation}
 The well known inequality
 \begin{equation}\label{eq:famousone}
-\ln(1-x)>x, 
\end{equation}
holding for $0<x<1$, implies
$-\ln (1-  \left(\frac12\right)^d\frac 14)\geq   \left(\frac12\right)^d\frac 14\geq \frac 1{16}$ from which we get the claimed lower bound on $-\ln(P_1+P_2)$ for $d\in \{1,2\}$.

\medskip
\noindent
Now let us consider the case $d\geq 3$. In this case we use Chernoff bound \cite{motrag}  to derive an upper bound on $P_1$.
Indeed, entries $M'(i,1),\ldots,M'(i,k)$ are $k$ i.i.d. Bernoulli random variables with probability of success equal to $p$, and
consequently, 
$\sum_{j\in M'}  M'(i,j)$ has binomial distribution
 with expectation $\mu=kp$.	
Chernoff bound  implies 
$Pr\{\sum_{j\in M'}  M'(i,j)> \mu(\delta+1)\}\leq \left({e^{\delta}\over(1+\delta)^{1+\delta}}\right)^{\mu}, \mbox{ for $\delta>0$}.$
By setting  $\mu=pk$ and $(\delta+1)\mu=d$, we obtain
$Pr\{\sum_{j\in M'}  M'(i,j)> d\}\leq e^{d-pk} \left({pk\over d}\right)^d$.
We can use this inequality to limit from above the second probability in the righthand side of (\ref{eq:P_1}), thus obtaining
\begin{eqnarray}\label{eq:P_1_2}
P_1&\leq &(1-p)^k+e^{d-pk} \left({pk\over d}\right)^d.
\end{eqnarray}
In order to derive an upper bound on $P_2$, we upper bound $(1-p)^{k}\sum_{i=1}^d{m-1\choose i} \left({p\over 1-p}\right)^{i}$ in (\ref{eq:P_2}) by 
\begin{equation}\label{eq:P_2new}
(1-p)^{k}\sum_{i=1}^{m-1}{m-1\choose i} \left({p\over1-p}\right)^{i}=(1-p)^{k}\left(\left(1+ {p\over1-p}\right)^{m-1}-1\right)
=(1-p)^{k-m+1}-(1-p)^k.
\end{equation}
Upper bound (\ref{eq:P_1_2})  on $P_1$ and upper bound (\ref{eq:P_2new}) on $P_2$
imply
\begin{equation}\label{eq:P1P2new2}
P_1+P_2\leq  e^{d-pk} \left({pk\over d}\right)^d+ (1-p)^{k-m+1}.
 \end{equation}
By setting   $p=\frac d{4k}$ in  (\ref{eq:P1P2new2}), we obtain 
\begin{equation}\label{eq:P1P2_1}
 P_1+P_2\leq e^{d-\frac d 4} \left({1\over 4}\right)^d+ \left(1-\frac d{4k}\right)^{k-m+1}. 
\end{equation}
We will prove that 
\begin{equation}\label{eq:firstcase}
e^{d-\frac d {4}} \left({1\over 4}\right)^d<  \frac 13\left(1-\frac d{4k}\right)^{k-m+1}.
\end{equation}
Inequality (\ref{eq:firstcase}) holds if and only if 
\begin{equation}\label{eq:firstcase1}
{d-\frac d {4}} -d\ln 4+\ln 3< (k-m+1)\ln  \left(1-\frac d{4k}\right).
\end{equation}

The following  well known inequality
 \begin{equation}\label{notfamousone}
-\ln(1-x)\leq {x\over 1-x}, \mbox{ holding for all $0\leq  x<1$},
\end{equation}
implies that $\ln(1-\frac d{4k})\geq -\frac d{4k-d}$, and consequently, the righthand side of (\ref{eq:firstcase1}) is larger than $(k-m+1)\left(-\frac d{4k-d}\right)$.
It follows that inequality (\ref{eq:firstcase1}) holds if 
\begin{equation}\label{eq:firstcase2}
{d-\frac d {4}} -d\ln4+\ln 3 < (k-m+1)\left(-\frac d{4k-d}\right),
\end{equation}
which is satisfied for 
\begin{equation}\label{eq:firstcase3}
\left(\frac {k-m+1}{4k-d}\right)< -\frac {3}{4} +\ln4-\frac{\ln 3}d.
\end{equation}
The above inequality is satisfied for $d\geq 3$. Indeed, it is $\left(\frac {k-m+1}{4k-d}\right)\leq \left(\frac {k-d+1}{4k-d}\right)\leq \left(\frac {k-2}{4k-3}\right)< \frac 14$, whereas 
$-\frac {3}{4} +\ln4-\frac{\ln 3}3>0.26$. Therefore, we have shown that inequality (\ref{eq:firstcase}) holds.
Inequalities (\ref{eq:P1P2_1}) and (\ref{eq:firstcase}) imply 
\begin{equation}\label{eq:P1P2_2}
 P_1+P_2\leq \frac 4 3 \left(1-\frac d{4k}\right)^{k-m+1}. 
\end{equation}
Therefore, we have that
\begin{equation}\label{app10}
 -\ln(P_1+P_2)\geq -\ln\left(\frac 4 3\right)-(k-m+1)\ln \left(1-\frac d{4k}\right). 
\end{equation}
By inequality (\ref{eq:famousone}), we have that 
$$-\ln\left( 1-\frac d{4k}\right)>\left(\frac d{4k}\right).$$
The above inequality along with inequality (\ref{app10}) implies the claimed lower bound on $-\ln(P_1+P_2)$ for $d\geq 3$.
\end{document}